# THE ANTHROPIC ARGUMENT AGAINST INFINITE PAST AND THE EDDINGTON-LEMAÎTRE UNIVERSE

Milan M. Ćirković
*Astronomical Observatory, Volgina 7, 11000 Belgrade, SERBIA*

**Abstract.** This study in the philosophy of cosmology is a part of an ongoing effort to investigate and reassess the importance of the anthropic (Davies-Tipler) argument against cosmologies containing the past temporal infinity. Obviously, the prime targets of this argument are cosmological models stationary on sufficiently large scale, the classical steady state model of Bondi, Gold and Hoyle being the best example. Here we investigate the extension of application of this argument to infinitely old non-stationary models and discuss additional constraints necessary to be imposed on such models for the edge of the anthropic argument to be preserved. An illustrative counterexample is the classical Eddington-Lemaître model, in the analysis of which major such constraints are presented. Consequences of such an approach for our understanding of the nature of time are briefly discussed.

## 1. Introduction: eternal universe and anthropic principle

The formulations of various anthropic principles (Dicke 1961; Carter 1974) have necessarily impacted both development of cosmological theory and its philosophical foundations. One of the interesting applications of anthropic principles in practical cosmological work is the general argument, conceived by Paul C. W. Davies, and developed by Frank Tipler, aimed at refuting specific cosmological models involving past temporal infinity. The first such cosmological models which come to mind are steady-state models, in particular the classical steady state model of Bondi and Gold (1948) and Hoyle (1948). However, it is not difficult to perceive that the domain of applicability of this argument is wider than the steady state models. This argument, apart from its other applications, has some interesting consequences for our notion of the cosmic time itself. In addition, it serves as a powerful counterexample to those criticisms of anthropic principles which effectively reject them as shallow and uninformative tautologies (Earman 1987). In this note, we shall discuss some of the basics of this argument, and particularly restriction of its field of applicability from all models with a past infinity to the sub-class of the latter possessing what we shall call non-trivial past infinity. The detailed discussion is relegated to the future work, currently in preparation.

The anthropic argument against cosmologies postulating eternally existing universe has been first mentioned by Davies in his brief critique of Ellis (1978; Ellis, Maartens and Nel 1978) cosmological model (see Davies 1978),

> There is also the curious problem of why, if the Universe is infinitely old and life is concentrated in our particular corner of the cosmos, it is not inhabited by technological communities of unlimited age.

As mentioned by Barrow & Tipler (1986) in their encyclopaedic monograph, this is historically the first instance in which an anthropic argument has been used against a cosmology containing past temporal infinity, and it is indeed fascinating that nobody had considered it before. The suprise is strengthened by the fact that such cosmologies in scientific or half-scientific form has existed since the dawn of the modern science. Ancient cosmologies postulating an eternal universe began with Empedocles of Acragas (VI century B.C.), whose periodically repeating universe has many uncanny resemblances to the modern oscillatory cosmological models (e.g. Guthrie 1969; O'Brien 1969; Diels 1983).

The anthropic argument against steady-state theories mentioned by Davies has been subsequently expanded and elaborated by Tipler (1982). In that work it has been shown that this argument applies to "all universes which do not change with time in the large", and particularly those which satisfy the Perfect Cosmological Principle (Bondi & Gold 1948). The discrete Markov chain reccurence of the type discussed by Ellis & Brundrit (1979) has also been used in the discussion of Tipler (1982), although, as we shall discuss in detail in a future study, its use is largely superfluous, since even much weaker hypothesis produces the same disastrous effects for the cosmologies with past temporal infinities.

The essence of Tipler's (1982) discussion is the circumstance that, given some usual symmetries of spacetime, for each event $p$, its past light cone intersects all world lines corresponding to history of an intelligent species. Thus, at least one out of infinite number of such species, could travel along the time-like geodesic to $p$ (or just send signals). Since $p$ may be any event, like our reading of Tipler (1982) paper, or any other occurence in the Solar system, it is completely unexpected that we are not already part of an intelligent community of an arbitrarily long age. Again, it is important to stress **non-exclusivity** of this argument: even if 99.99% (or indeed any fraction less than unity) of intelligent communities arising at, say, $q$ would not expand further than some limited neighbourhood $q + \varepsilon$, in an infinitely old universe there would still be at least one intelligent community at any point $p$ in spacetime, no matter how big $|p - q| / \varepsilon$ is.

But Tipler (1982) goes further and claims that,

> Since all possible evolutionary sequences have occured to the past of $p$, one of these evolutionary sequences consists of the random assembly, without assistance of any intelligent species whatsoever, of a von Neumann probe out of the atoms of interstellar space. Such a random assembly would occur an infinite number of times to the past of $p$, by homogeneity and stationarity in an infinite universe. At least one of these randomly assembled probes would have the motivations of a living being, that is to expand and reproduce without limit.

This scenario, although not at all fantastic as it may seem at first, raises several questions still lacking elaboration. How could we possibly know that the set of all "favorable" spontaneously assemled von Neumann probes is of non-zero measure in the set of all possible spontaneously assembled probes. the question of motivation, which is not so easily quantifiable, becomes crucial here. For instance, why not postulate an assembly of von Neumann probe designed to search and destroy other von Neumann probes? What is the relative weight of colonizing (vs. destructive, altruistic, etc.) motivation, and how can one determine it? This motivation problem is avoided if we stick to more restrictive requirement that only communities of evolved intelligent beings create such probes (i.e. create them at timescales many orders of magnitude shorter than those required for spontaneous assembly Tipler describes).

While one may argue that motivation is necessarily linked to the level of complexity, and therefore one expects the spontaneously assembled self-reproducing automata will have basically the same motivations we perceive in biological systems on Earth (Tipler, private communication), this issue is not clear at all. Therefore, we shall use the weaker version of Davies, requiring only that conditions favorable for evolutionary emergence of intelligent communities similar to ours persist. Such communities, in this version of the argument, present the source of technologization we fail to observe.

## 2. A counterexample: a universe with false infinity

For the sake of better understanding of the issues involved, let us consider a counterexample of a cosmological model involving past temporal infinity which the Davies-Tipler argument does not apply to. This is the Lemaître-Eddington universe, which was quite popular in the 1925-1935 period. Good description of this model can be found in the classical Bondi's textbook on cosmology (Bondi 1961). Having appeared on the cosmological scene after the realization of instability of original Einstein static universe (Einstein 1917), this model

> ...has therefore an infinite past which was spent in the Einstein state. This has greatly attracted investigators since it seemingly permits an arbitrarily long timescale of evolution. The picture of the history of the universe derived from this model, then, was that for an infinite period in the distant past there was a completely homogeneous distribution of matter in equilibrium in the Einstein state until some event started off the expansion, which has been going on at an increasing pace ever since. the condensation of the galaxies and the stars from the primeval matter took place at the time the expansion began, but this development was stopped later by the decrease of average density due to the progress of the expansion.[1]

From the formal point of view, in accordance with the Weil postulate, the Eddington-Lemaître universe has an infinite past, i.e. the initial state is given by the formal limit $t \rightarrow -\infty$. This state is characterized, as in the original Einstein model, by detailed equilibrium between attractive force of gravity and repulsive force due to the positive cosmological constant. In addition, such state is expected to achieve perfect thermodynamical equilibrium. However, this is a "false" infinity, at least in the context of anthropic reasoning, because the period of time in which there are conditions enabling creation of intelligent observers is necessarily finite. In addition, this period is approximately equal to the time past since the beginning of the expansion. The period of complete homogeneity can be regarded as a state analogous to the epochs of complete dominance of Love or Strife in the cosmology of Empedocles (O'Brien 1969). In both cosmologies it is necessary to invoke a state which prevents propagation of information from arbitrarily distant past to the present epoch. In both cases this goal is achieved by postulating states with sufficiently high degree of symmetry. Obviously, in the case of the Eddington-Lemaître universe, the anthropic argument is inapplicable, since the effective past is finite. Intelligent observers (as well as spontaneously assembled von Neumann probes!) possess only a finite time for technologization of their cosmic environment. This is valid for the generic version of the Eddington-Lemaître model. Of course, the model pretending to describe the real universe is normalized to the present expansion rate, and therefore we conclude that this effective age is similar to the age of galaxies, or again of the order of $H^{-1}$. Therefore, the incompatibility argument in the core of the Davies-Tipler argument is lost and reduces to much weaker Fermi "paradox", as we shall see in the further discussion.

---
[1] Bondi (1961), p. 118.

Probably the more physical and meaningful way of restating the entire situation is to reject the notion of infinite age of the Eddington-Lemaître model as a hollow formalism. So-called Aristotle principle tells us that there is no time without changeable world. The state of perfect equilibrium in Eddington-Lemaître model in the $t \rightarrow -\infty$ limit is exactly such unchangeable state, without means of determining either direction or the rate of passage of time. In the sense of modal version of the Aristotle principle, the temporal infinity in this model thus collapses into a purely formal notion. Newton-Smith's formulation of this principle

> There is a period of time between the events $E_1$ and $E_2$ if and only if relative to these events *it is possible* for some event or events to occur between them,

(Newton-Smith 1980, p. 44) explicitly points out to indistinguishability of moments in the state of complete thermodynamical equilibrium (see also Arsenijević 1986). The same applies to the far future of the universe in which, according to many models, the state of heat death may occur. Barrow & Tipler (1978), in one of the first papers devoted to the cosmological future, suggest that formally infinite future should be substituted with a finite interval, through an appropriate coordinate transformation. A sort of counterexample, confirming the general thesis that the cosmic time established by the Weil postulate should not be regarded as sacrosanct, is the diverging number of (possible) events in the finite temporal vicinity of either initial or the final global singularity. In such a situation a finite cosmic time may be less appropriate than an alternative infinite timescale (e.g. Misner 1969).

In conclusion, the past temporal infinity in the Eddington-Lemaître model is **trivial** from the anthropic point of view, and the Davies-Tipler argument is inapplicable. Thus, one should reduce the realm of applicability of the latter argument to cosmological models containing non-trivial past infinities. The residual problem in each case is what is traditionally called Fermi paradox. It is clear, for instance, that the same applies to classical oscillatory universes (Tolman 1934), which present the modern rehash of the Empedocles' cosmology. Examples of such non-trivial past infinities are those inclusive in any sort of static model, or stationary model incorporating any highly symmetrical Copernican postulate (of which the most famous is the Bondi-Gold perfect cosmological principle).

**3. Entropy and arrow of time in light of the anthropic argument**

It should be immediately noted that the Davies-Tipler argument as exposed above is different from the unlimited entropy argument usually used against cosmologies with past infinities (although the two are related, as we shall see below): why haven't irreversible processes, in accordance with the thermodynamical laws, generated infinite amount of entropy in the universe by now? Davies himself used the same argument against the Hoyle-Narlikar cosmology in his review of the latter in "Nature" (Davies 1975), and Tipler (1982) mentions it in somewhat restricted sense, as the Olbers' paradox (again, expanded discussion may be found in Barrow & Tipler 1986). The classical steady-state theory alleviates this problem by continuous creation of matter, and additional assumption that newly created matter is in low-entropy state. But cosmologies excluding creation of matter (such as, for instance, Einstein original static universe, or Hoyle-Narlikar conformally invariant cosmology[2]) are faced with this argument in a very serious form. Still, this thermodynamical argument against steady-state models is qualitatively different from the Davies-Tipler argument we are dealing here, although both show how difficulties arise when currently observable processes are extrapolated backward in the past eternity. The latter argument is based, essentially, on the diametrically opposed process: growth of complexity, which results in emergence of technological communities at some finite time. In the former case, we perceive increase in entropy in laboratory experiments; in the latter case, we perceive our laboratories themselves, and—in a sense—the very results of our former observations.

However, if we accept the notion that thermodynamical arrow of time is essentially a product of the cosmological initial conditions [as has been suggested already by Boltzmann, and the best modern treatment can be found in the nice book of Price (1996)], the conditions in the $t \rightarrow -\infty$ limit obtain a new and profound significance. In any attempt to build an atemporal ("tenseless") picture of the universe, such as Price's, some sort of symmetry between the initial and the final conditions has to be satisfied. If we are restricted to Friedmann universes, we encounter severe problems, if not ready, like Gold (1962) to insist on a highly special, time-symmetric universe. These problems become even more severe for model universes like the one of Ellis and coworkers in which there is no large scale motion of matter and energy, and the cosmological arrow of time is completely lost. The alternative solution, the idea that asymmetric physical *laws* are necessary for cosmology, has been with us for very long time, actually since 1930-ies, from early work of Russian cosmologist Matvei Bronstein. Before he was brutally murdered by Soviet socialists in 1938, he presciently wrote in the paper published in 1933 that physical theory upon which the cosmological solutions can be based cannot be symmetrical with respect to the interchange of the past and the future (Bronstein 1933). This view has been vigorously put forward by Rodger Penrose in recent years (Penrose 1979, 1989), but actually it underlies most of the practical cosmological work, **in particular** since the victory of big bang models over their great steady-state rival (Kragh 1996). In this manner, the anthropic argument helps us further highlight the implausibility of stationary cosmologies, in which the different treatment of the arrow of time would be required in order to avoid all imaginable counterfactual consequences of its dissolution.

**Acknowledgement.** The author is happy to acknowledge his friends and teachers in the Alternative Academic Network, Prof. Petar Grujić and Prof. Miloš Arsenijević, for many useful discussions on the subject and general encouragement. The kind support of Prof. Fred C. Adams, Prof. Helge Kragh, Mr. Mašan Bogdanovski, Mr. Srđan Samurović and Vesna Milošević-Zdjelar is also acknowledged.

---

[2] There is a slight confusion in the literature which of several different cosmological models is correctly called Hoyle-Narlikar cosmology. Here, we attach this name only to the conformally invariant model with conserved number of particles and variable masses, such as exposed in Hoyle & Narlikar (1972) and Hoyle (1975) papers.